

\input harvmac.tex
\noblackbox
\lref\Klemm{A. Klemm and R. Schimmrigk, ``Landau-Ginzburg String Vacua,''
{\it Nucl. Phys.} {\bf B411} (1994) 559.}
\lref\GSW{M. Green, J. Schwarz, and E. Witten, ``Superstring Theory: Volume
2,'' Cambridge University Press, 1987.}
\lref\Slansky{R. Slansky, ``Group Theory for Unified Model Building,''
{\it Phys. Rept.} {\bf 79} (1981) 1.}
\lref\Wilmod{G. Athanasiu, J. Atick, M. Dine, and W. Fischler,
``Remarks on Wilson Lines, Modular Invariance, and Possible String
Relics in Calabi-Yau Compactifications,'' {\it Phys. Lett.}
{\bf B214} (1988) 55.}
\lref\Schimm{R. Schimmrigk, ``A New Construction of a Three Generation
Calabi-Yau Manifold,'' {\it Phys. Lett.} {\bf 193B} (1987) 175.}
\lref\dealwis{S.P. DeAlwis, ``Gauge Symmetry Breaking in
Compactified Heterotic Strings,'' {\it Phys. Lett.} {\bf B216}
(1989) 277.}
\lref\Nath{R. Arnowitt and P. Nath, ``Predictions from Three
Generation Calabi-Yau String Theory,'' {\it Phys. Rev.} {\bf D42}
(1990) 2948.}
\lref\Segre{J.D. Breit, B. Ovrut, and G. Segre, ``$E_{6}$ Symmetry
Breaking in the Superstring Theory,'' {\it Phys. Lett.} {\bf 158B}
(1985) 33.}
\lref\witold{E. Witten, ``Symmetry Breaking Patterns in Superstring
Models,'' {\it Nucl. Phys.} {\bf B258} (1985) 75.}
\lref\Hubsch{T. Hubsch, {\it Calabi-Yau Manifolds: A Bestiary for
Physicists}, World Scientific, 1992.}
\lref\threeform{P. Candelas, ``Yukawa Couplings Between (2,1) Forms,''
{\it Nucl. Phys.} {\bf B298} (1988) 458.}
\lref\Vafa{C. Vafa, ``Modular Invariance and Discrete Torsion on
Orbifolds,'' {\it Nucl. Phys.} {\bf B273} (1986) 592.}
\lref\ThreeCY{P. Candelas, M. Lynker, and R. Schimmrigk,
``Calabi-Yau Manifolds in Weighted P4,'' {\it Nucl. Phys.} {\bf B341}
(1990) 383.}
\lref\Tororb{See e.g.
G. Aldazabal, A. Font, L. Ibanez, and A. Uranga, ``String GUTS,''
hep-th/9410206;
A. Font, L. Ibanez, F. Quevedo, and
A. Sierra, ``The Construction of `Realistic' Four-Dimensional
Strings through Orbifolds,''
{\it Nucl. Phys.} {\bf B331}
(1990) 421; and references therein.}
\lref\DK{J. Distler and S. Kachru, ``(0,2) Landau-Ginzburg Theory,''
{\it Nucl. Phys.} {\bf B413} (1994) 213, hep-th/9309110.}
\lref\DKtwo{J. Distler and S. Kachru, ``Singlet Couplings and (0,2)
Models,'' {\it Nucl. Phys.} {\bf B430} (1994) 13, hep-th/9406090.}
\lref\DKthree{J. Distler and S. Kachru, ``Quantum Symmetries and
Stringy Instantons,'' {\it Phys. Lett.} {\bf B336} (1994) 368,
hep-th/9406091.}
\lref\NewIss{E. Witten, ``New Issues in Manifolds of SU(3) Holonomy,''
{\it Nucl. Phys.} {\bf B268} (1986) 79.}
\lref\WitSilvtwo{E. Silverstein and E. Witten, to appear.}
\lref\Dixon{T. Banks, L. Dixon, D. Friedan, and E. Martinec, Mixing
Angle.}
\lref\CHSW{P. Candelas, G. Horowitz, A. Strominger, and E. Witten,
``Vacuum Configurations for Superstrings,'' {\it Nucl. Phys.} {\bf B258}
(1985) 46.}
\lref\DistGr{J. Distler and B. Greene, ``Aspects of (2,0) String
Compactifications,'' {\it Nucl. Phys.} {\bf B304} (1988) 1.}
\lref\DSWW{M. Dine, N. Seiberg, X.G. Wen and E. Witten,
``Non-Perturbative Effects on the String World Sheet I,''
{\it Nucl. Phys.} {\bf B278} (1986) 769, ``Non-Perturbative Effects on
the String World Sheet II,'' {\it Nucl. Phys.} {\bf B289} (1987) 319.}
\lref\GRoss{B.R. Greene, ``Special Points in Three-Generation Moduli
Space,'' {\it Phys. Rev.} {\bf D40} (1989) 1645;
B.R. Greene, K. Kirklin, P. Miron, and G.G. Ross, ``A Three Generation
Superstring Model II: Symmetry Breaking and the Low Energy Theory,''
{\it Nucl. Phys.} {\bf B292} (1987) 606, ``A Three Generation Superstring
Model I: Compactification and Discrete Symmetries,'' {\it Nucl. Phys.}
{\bf B278} (1986) 667.}
\lref\FFermion{See e.g. S. Chaudhuri, S. Chung, G. Hockney, and
J. Lykken, ``String Consistency for Unified Model Building,''
hep-ph/9501361; G. Cleaver, ``GUTS with Adjoint Higgs Fields
from Superstrings,'' hep-th/9409096;
A. Faraggi, ``$Z_{2} \times Z_{2}$ Orbifold Compactification -- The Origin
of Realistic Free-Fermionic Models,''
hep-ph/9501288; and references therein.}
\lref\Scheich{C. Scheich and M. Schmidt, ``Phenomenological Discussion of
the Three Generation Gepner Model,'' {\it Int. Jour. Mod. Phys.}
{\bf A7} (1992) 8021.}
\lref\Gepner{D. Gepner, ``String Theory on Calabi-Yau Manifolds:
The Three Generations Case,'' Princeton preprint PUPT-88/0085.}
\lref\threshold{L. Ibanez, ``Gauge Coupling Unification: Strings
versus SUSY GUTs,'' {\it Phys. Lett.} {\bf B318} (1993) 73, hep-ph/9308365;
L. Ibanez, D. Lust, and G. Ross, ``Gauge Coupling Running in
Minimal $SU(3)\times SU(2)\times U(1)$ Superstring Unification,''
{\it Phys. Lett.} {\bf B272} (1991) 251, hep-th/9109053.}

\Title{\vbox{\hbox{HUTP-95/A003}\hbox{\tt hep-th/9501131}}}
{Some Three Generation (0,2) Calabi-Yau Models}
\bigskip
\centerline{Shamit Kachru\footnote{$^\dagger$}
{Junior Fellow, Harvard Society of Fellows}}
\bigskip\centerline{\it Lyman Laboratory of Physics}
\centerline{\it Harvard University}\centerline{\it Cambridge, MA 02138}

\vskip .3in

It has recently been realized that a large class of Calabi-Yau models
in which the VEV of the gauge connection is not set equal to the spin
connection of the Calabi-Yau manifold are
valid classical solutions of string theory.  We provide some
examples of three generation models based on such generalized
Calabi-Yau compactifications, including models with observable
gauge group $SU(3)\times SU(2)\times U(1)$.

\Date{\it {January 1995}} 


\newsec{Introduction}

String theory is the leading (and at present, only) candidate for a
unified theory of the fundamental interactions.  However, daunting
challenges remain to be overcome, both in our fundamental understanding
of strings and in connecting string theory with low energy physics.

Although at present we cannot make definite statements
about generic properties of string-based phenomenology, one of the obvious
things we can do is construct examples of pseudo-realistic string
compactifications and explore, with judicious assumptions, the resulting
phenomenology.  The purpose of such explorations is of course not to
find $\it the$ model of nature, but to see if one can find generic
advantages or deficiencies of string-derived models.

To that end, in this paper we provide several examples of three-generation
string compactifications on Calabi-Yau manifolds \CHSW.
While the Calabi-Yau
models most commonly studied to date have been (2,2) supersymmetric models
which naturally yield an $E_{6}$ effective GUT group, (0,2) Calabi-Yau
models provide a much broader class of compactifications in which one can
also naturally obtain $SO(10)$ or $SU(5)$ as the effective gauge
group \refs{\NewIss,\DistGr}.  It has recently been proved that, contrary to
previous expectations \DSWW, many (0,2) Calabi-Yau models
are true solutions of string theory (both non-perturbatively
in the sigma model expansion and to all orders in the string loop expansion)
\refs{\DKtwo,\WitSilvtwo}.\foot{The conclusions of the
forthcoming paper \WitSilvtwo\ in particular
are stronger than those of \DKtwo\
and imply that
all of the models
considered here are $\it bona~ fide$ solutions of string theory.}
It therefore seems worthwhile to explore the enhanced phenomenological
opportunities in this large new class of classical string vacua,
and we take a first step in that direction in this paper.

In \S2, we briefly describe the class of models under consideration \DK.
In \S3, we provide four examples in this class of models which
yield three generations of chiral fermions in the normal unification
representations of $E_{6}$ and $SU(5)$.   These models
are $\it not$ related to the (2,2) theories on the same Calabi-Yau
manifolds by holomorphic deformation.  Two of the examples, with
gauge group $SU(5)$, are on manifolds with nontrivial $\pi_{1}$.
In \S4,
using Wilson lines as a tool
for gauge symmetry breaking
\refs{\witold,\Segre}, we break the gauge group in
these two $SU(5)$ examples
to $SU(3)\times SU(2)\times U(1)$.
It will be clear
that our results are in no sense exhaustive -- one could certainly
construct many other models along similar lines.

To our best knowledge, only one three-generation Calabi-Yau model has been
explored in any depth in the literature -- see
\refs{\GRoss,\Nath,\Scheich,\Gepner,\Schimm}\
and references therein.  Several three generation (2,2) models have been
listed in \ThreeCY\ and \Klemm\ (the latter
reference classifies all CY hypersurfaces in
weighted projective four-spaces), all of which are compactifications
on manifolds with trivial $\pi_{1}$.\foot{I am informed by A. Klemm
that it is extremely difficult if not impossible
to find examples of three generation
models with nontrivial $\pi_{1}$ by orbifolding higher generation models
in this class; this has been investigated by A. Niemeyer.}
Considerably more work has been directed towards exploring semi-realistic
string models based on toroidal
orbifolds \Tororb\ and free-fermions \FFermion.  It may
prove interesting to investigate
the phenomenology of
the models presented here, or other similar models; we hope to
undertake such investigations in the near future.

\newsec{A Class of (0,2) Calabi-Yau Models}

Recall that the data which enters in specifying a (0,2) Calabi-Yau model
is a choice of Calabi-Yau manifold $M$ and stable, holomorphic vector
bundles $V_{1}$ and $V_{2}$ (representing the vacuum configurations of
the gauge fields in the observable
and hidden $E_{8}$ of the heterotic string)
satisfying
\eqn\ctwo{c_{2}(M) ~=~ c_{2}(V_{1}) ~+~c_{2}(V_{2})}
\eqn\cone{c_{1}(V_{1,2}) ~=~0~{\rm mod}~2~.}
Here the $c_{i}$ are the Chern classes of the vector bundles in
question.  \ctwo\ is the well known anomaly cancellation condition,
while \cone\ is the requirement that $V_{1}$ and $V_{2}$ admit
spinors.

In addition to these topological conditions, there
are  perturbative conditions
for conformal invariance of a (0,2) Calabi-Yau $\sigma$-model \NewIss.
At lowest order, we
must require that the metric on $M$ be the familiar Ricci-flat Kahler
metric
$g_{i\bar j}$ whose
existence is guaranteed by Yau's theorem.
In addition, the connections on $V_{1,2}$ must satisfy
\eqn\conn{g^{i\bar j}F_{i\bar j} ~=~ 0~.}
For stable bundles $V_{1,2}$, a theorem of Uhlenbeck and Yau guarantees
the existence of a solution to \conn\ as long as the integrability
condition
\eqn\integ{\int_{M} J\wedge J\wedge c_{1}(V_{1,2}) ~=~0}
is met, where $J$ is the Kahler form of $M$.
We will satisfy \integ\ by choosing $c_{1}(V_{1,2}) =0$ (as indeed we
must if we work on a manifold $M$ with $h_{1,1}=1$).
Higher orders of sigma model perturbation theory do not lead to
any further conditions on $M$ or $V_{1,2}$.

We will confine our attention to non-singular Calabi-Yau manifolds
$M$, defined
by the vanishing loci of $N$ equations
of degree $d_{i}$ ($1\leq i \leq N$) in some $WCP^{N+3}_{w_{1}....w_{N+4}}$
with homogeneous coordinates $\phi_{j}, ~1\leq j \leq N+4$.
In such a model $M$ has only one harmonic (1,1) form $J$ (it inherits the
Kahler class of
the ambient projective space), and one finds that
\eqn\intKah{\int_{M} ~J\wedge J\wedge J ~=~{{\Pi~ d_{i}}\over {\Pi w_{j}}}~.}
This will be useful momentarily.\foot{Formulas such as this and other
basic wisdom concerning Calabi-Yau spaces can be found in \Hubsch.}

In the models of interest with
$V_{1}$ of rank 3, 4, or 5 (spacetime gauge group
$E_{6}$, $SO(10)$, or $SU(5)$), the net number
of generations ($\#$ generations -  $\#$ antigenerations)
$N_{gen}$ of chiral fermions in the observable sector
(in the $\bf 27$ of $E_{6}$, $\bf 16$ of $SO(10)$ or
$\bf 10 + \bf \bar 5$ of $SU(5)$) is given by
\eqn\gen{N_{gen} ~=~ {1 \over 2} \vert \int_{M} ~c_{3}(V_{1}) \vert~.}
\gen\ reflects the fact that massless fermions in four dimensions
correspond to zero modes of the Dirac operator on the compactification
manifold $M$, and follows from the index theorem.
It reduces to the familiar expression ${1\over 2}|\chi|$ in the (2,2)
case.  To determine separately the
number of generations and the number of anti-generations
instead of just the difference, one must also compute the dimension
of $H^{1}(M,V_{1})$.

Following \DK, we consider vacuum gauge bundles which are
defined by the following exact sequence:
\eqn\bundle{0\to V\to \bigoplus_{a=1}^{r+M} \CO(n_a)  {\buildrel
{\otimes F_a^{i}(\phi)}\over{\hbox to 30pt{\rightarrowfill}}}
\bigoplus_{i=1}^{M} \CO(m_{i})\to 0~.}
Here
$r=3,4,5$ yields gauge group $E_{6}$, $SO(10)$ or $SU(5)$,
and the $n_{a}$ and $m_{i}$ are
positive integers with $\sum m_{i}=\sum n_{a}$ (guaranteeing
that $c_{1}(V)=0$).
$\CO(a)$ denotes the $a$th power
of the hyperplane bundle of the ambient weighted
projective space, and the $F_{a}^{i}(\phi)$
are polynomials homogeneous of degree $m_{i}-n_{a}$ in the $\phi$s which
never simultaneously vanish on $M$.
For shorthand, we use the notation $V = ( \{ m_{i} \}; \{ n_{a} \} )$.
The worldsheet quantum field theory which describes this spacetime
model can be obtained as the infrared limit of a linear
sigma model with $U(1)$ gauge group (for more details the reader
should consult \DK) and
the linear sigma model will contain $r+M$ left-moving fermions
$\lambda_{1},\cdots,\lambda_{r+M}$ with gauge charges $(n_{1},\cdots,
n_{r+M})$.

In general we will have nontrivial factors embedded in both the
observable and hidden $E_{8}$s, and we will denote them by
$V_{1}$ and $V_{2}$.
For a theory with $V_{1}$ described as in \bundle, one finds that
\eqn\cthree{c_{3}(V_{1}) ~=~-{1\over 3} ( \sum_{i} m_{i}^{3} ~-~
\sum_{a} n_{a}^{3}) ~J^{3}~.}
Combining \cthree\ with \gen\ and \intKah, we see that determining the
net number of
generations of chiral fermions in such a compactification is a simple exercise
in arithmetic.  $H^{1}(M,V_{1})$ also has a convenient deformation theoretic
representation as M-tuples $(P_{1}(\phi),\cdots,P_{M}(\phi))$ modulo
$\{ (F_a^{1}(\phi),\cdots,F_{a}^{M}(\phi)) \}$ where of course $P_{i}$ has
degree $m_{i}$.  This allows us to compute dim $H^{1}(M,V_{1})$ in the
examples of \S3 and verify that not only is \gen\ equal to three
but there are in fact precisely three generations and no antigenerations.

\newsec {Some Three Generation Compactifications}

We now use the apparatus of \S2 to construct some three-generation models.
We begin with two very simple examples of $E_{6}$ theories and then
discuss two more involved examples, which both give rise to $SU(5)$
gauge group (and are on non-simply connected manifolds, which admit
gauge symmetry breaking by
Wilson lines).  In the second $SU(5)$ example, the hidden sector $E_{8}$
remains unbroken. In \S4, we break the gauge group of the two $SU(5)$
examples down to $SU(3)\times SU(2)\times U(1)$ by using Wilson lines.

\subsec {Example 1: An $E_{6}$ Model}

Consider the compactification on the Calabi-Yau
hypersurface $M$ defined by the vanishing locus of a degree ten
polynomial in $WCP^{4}_{1,1,1,2,5}$ with
$V_{1} = (2,2,2;1,1,1,1,1,1)$ and
$V_{2} = (9;1,1,2,2,3)$.  One easily computes that
$\int_{M} J^{3} = 1$ for this manifold, so $\int_{M} c_{3}(V_{1}) = -6$.
Hence, this model has a net of three generations in the observable sector.
The effective GUT group is $E_{6}$, and the charged matter
fields transform in the $\bf 27$ of $E_{6}$.  Using the ideas in
\S2 it is easy to verify that
dim $H^{1}(M,V_{1}) = 3$ so we have three generations and no
antigenerations.   This is true in our other examples as well,
so we will not repeat the statement on each occasion.

\subsec {Example 2: Another $E_{6}$ Model}

Now, we look at an example on a complete intersection manifold.  Let $M$
be the intersection of the vanishing loci of two degree six equations in
$WCP^{5}_{1,1,2,2,3,3}$.  Let $V_{1} = (2,2,2;1,1,1,1,1,1)$ and
$V_{2} = (7;1,1,1,2,2)$.  Then $\int_{M} ~c_{3}(V_{1}) = -6$, so once
again we have a compactification with three generations in the $\bf 27$
of $E_{6}$.

\subsec {Example 3: An $SU(5)$ Model}

Next, we find an $SU(5)$ theory on a manifold with $\pi_{1} = Z_{3}$.
Start as in Example 2, with $M$ being
the complete intersection of two degree six polynomials
$P_{1}$ and $P_{2}$ in $WCP^{5}_{1,1,2,2,3,3}$.
Choose $V_{1} = (3,3,2;1,1,1,1,1,1,1,1)$ and
$V_{2} = (6,3;1,1,1,2,2,2)$.
One can easily check that this theory has 9 generations in the
$\bf \bar 5 + \bf 10$ of $SU(5)$
in the observable sector, so we want to orbifold
by a freely acting $Z_{3}$ to obtain a three-generation model.

Consider the $Z_{3}$ action generated by $g$ which acts as follows
\eqn\gac{g:~~\phi_{1,3,5} \rightarrow \alpha\phi_{1,3,5},~~
\phi_{2,4,6}\rightarrow \alpha^{2}\phi_{2,4,6}~}
where $\alpha = e^{i{2\pi\over 3}}$.
Given these transformation laws, one can write down nonsingular choices of
$P_{1}$, $P_{2}$ which admit \gac\ as a symmetry; choose such an $M$.
Then not only is \gac\ a symmetry of this manifold, but for
generic choices it acts freely on $M$.

It must also preserve the holomorphic three-form $\Omega$ on $M$ if
the quotient is to be a solution of string theory.
By the general argument presented on p.495 of \GSW, one
knows that the holomorphic three-form of $M$ will always descend to the
quotient of $M$ by a freely acting symmetry. This
follows from the fact that the arithmetic genus
$\sum_{k=0}^{3} (-1)^{k} h^{0,k}$
of the quotient must
vanish.
Nonetheless,
it is instructive to explicitly verify the invariance of $\Omega$ (as
practice for the case of quotients by non-freely acting symmetries, for
example).
Following
\S3 of \threeform\
we see that the three-form can be written
\eqn\form{\Omega ~=~\oint \oint {{~\epsilon_{ijklmn} ~\phi_{i}~ d\phi_{j}\wedge
d\phi_{k} \wedge d\phi_{l} \wedge d\phi_{m} \wedge
d\phi_{n}}\over {P_{1}P_{2}}}}
where the contour integrals are taken about the
loci $P_{1}=0$ and $P_{2}=0$.
Since the numerator of the integrand transforms
with an $\alpha^{9} = 1$
while the denominator $P_{1}P_{2}$ is invariant as well, $\Omega$ does
descend to the $Z_{3}$ orbifold .

At this point we have seen that the $Z_{3}$ acts freely on $M$ and
preserves the holomorphic three-form, so if we were studying
a (2,2) model we would be done.  However, to take a quotient
of a (0,2) model by such a discrete symmetry group $G$, one must also
check that $G$ lifts to an automorphism of the vacuum gauge bundle
$V$.
In addition one has to check the level-matching conditions of \Vafa\ which
are necessary for modular invariance.   These conditions are more or
less automatically satisfied in (2,2) models but not (0,2) models.

A brief summary of these conditions is as follows:  Suppose
we wish to orbifold our (0,2) model by a $Z_{N}$ and let
$\beta = e^{{2\pi i}\over N}$.  Assume the $\phi_{i}$
transform as $\beta^{r_{i}}$, the fermions $\lambda^{1}_{a}$
associated with $V_{1}$ transform as $\beta^{r_{a}}$
and the fermions $\lambda^{2}_{b}$ associated with $V_{2}$
transform as $\beta^{\tilde r_{b}}$.  Then the conditions which
must be satisfied for $N$ even are
\eqn\neven{\sum_{i} (r_{i})^{2} ~=~\sum_{a} (r_{a})^{2} + \sum_{b}
(\tilde r_{b})^{2}~~{\rm mod} ~2N}
\eqn\neventwo{\sum_{i} r_{i} ~=~\sum_{a} r_{a} ~=~\sum_{b} \tilde r_{b}
= 0~~{\rm mod} ~2~.}
For odd $N$ one gets only the
analogue of \neven, and it must hold mod $N$ instead of
mod $2N$.  Note that these conditions are $\it necessary~$ for
modular invariance but are only known to be sufficient in the
case of free field theory (on
the worldsheet).  Indeed, in interacting (0,2) theories
there are indications that extra constraints may be needed to
ensure consistency at the one-loop level \DKthree.  We
will have nothing more to
say about this here, however.

Let us choose the extension of $g$ as follows.  Let $g$
act on the eight fermions associated to $V_{1}$ as
\eqn\gvone{g ~{\rm on}~ V_{1}:~~1, 1, 1, \alpha, \alpha,
\alpha^{2},\alpha^{2},\alpha^{2}~.}
Let $g$ act on the fermions of worldsheet $U(1)$ gauge charges $(1,1,1,2,2,2)$
associated with $V_{2}$ as
\eqn\gvtwo{g~{\rm on}~ V_{2}:~~1,\alpha,\alpha^{2},1,\alpha,\alpha^{2}~.}
It is now easy to check that the level-matching condition is satisfied.
Choose the $F_{a}^{i}$
involved in defining $V_{1}$ to insure that the combinations
$\lambda_{a}^{1} F_{a}^{i}$ (no sum on $a$) are $g$ invariant
(one can make such choices), and do the same for $V_{2}$.
This restricted set of $F$s represents the set for which the
chosen $g$ action is indeed an automorphism of $V$.

Having fulfilled
the various consistency conditions, we
see that we have found a freely acting $Z_{3}$
orbifold of the 9 generation theory in $WCP^{5}_{1,1,2,2,3,3}$.
The result
is a model with
three $\bf \bar 5 + \bf 10$s of $SU(5)$ as the charged observable
sector matter content.  In addition, the target manifold has a nontrivial
fundamental group $\pi_{1} = Z_{3}$.

It is worth emphasizing that
because we have obtained this model as the quotient of
the 9 generation model by a freely acting $Z_{3}$, the $\it massless$
states in the 3 generation theory simply correspond to the $Z_{3}$
invariant massless states in the 9 generation theory.
The twisted sectors only contribute massive states.
In the case of quotients by non-free group actions, one would in general
have some massless states coming from twisted
sectors as well.

\subsec{Example 4: Another $SU(5)$ Model}

The last example we provide is an $SU(5)$ theory on the $Z_{5}\times Z_{5}$
orbifold of the quintic first studied in \CHSW.  That is, we
choose for $M$ the Fermat quintic
\eqn\quintic{\sum_{i=1}^{5} \phi_{i}^{5}~=~0}
in $CP^{4}$.
Instead of starting with the (2,2) theory on this quintic, however, we
begin with $V_{1} = (3,3,3,3,3;1,1,1,1,1,2,2,2,2,2)$ and we leave
the second $E_{8}$ unbroken.  It is easy to check that
$\int_{M}~c_{3}(V_{1}) = -150$ so orbifolding by a freely acting
$Z_{5}\times Z_{5}$ will yield a three generation $SU(5)$ theory.

The $Z_{5}\times Z_{5}$ symmetry group of \quintic\ that we wish to
orbifold by is generated by
\eqn\gone{g_{1}:~~(\phi_{1},\phi_{2},\phi_{3},\phi_{4},\phi_{5})
\rightarrow (\phi_{2},\phi_{3},\phi_{4},\phi_{5},\phi_{1})}
\eqn\gtwo{g_{2}:~~\phi_{i} \rightarrow \alpha^{i}\phi_{i}~,}
where $\alpha = e^{{2\pi i}\over 5}$.
As discussed in \CHSW, this discrete symmetry group does act freely
on \quintic\ and hence preserves the holomorphic three-form.

Now, we need to assign
transformation laws to the $\lambda$s and
choose the holomorphic structure
of $V_{1}$ so that the symmetries \gone\ and \gtwo\
lift to automorphisms of $V_{1}$.
Let us imagine we are orbifolding first by \gone, then by \gtwo.
Denote by $\lambda_{1,\cdots,5}$ the five left-moving fermions of
worldsheet $U(1)$ gauge charge one
and by $\lambda_{6,\cdots,10}$ the five
of gauge charge two.  Choose
\eqn\sectone{F_{1}=(\phi_{1}^{2} + \phi_{3}\phi_{4}, \phi_{1}^{2},
\phi_{1}^{2}, \phi_{1}^{2}, \phi_{1}^{2})}
(the $i$th component of $F_{1}$ above represents $F_{1}^{i}$
in \bundle) and
similarly let $F_{2},\cdots,F_{5}$ be given by
\eqn\sectagain{F_{2}=(\phi_{2}^{2},\phi_{2}^{2} + \phi_{4}\phi_{5},
\phi_{2}^{2},\phi_{2}^{2},\phi_{2}^{2}),
 \cdots,F_{5} = (\phi_{5}^{2},\phi_{5}^{2},
\phi_{5}^{2},\phi_{5}^{2},\phi_{5}^{2} + \phi_{2}\phi_{3})~.}
To complete the specification of the holomorphic structure of $V_{1}$,
choose
\eqn\secttwo{F_{6}=(\phi_{1},0,0,0,0), F_{7}=(0,\phi_{2},0,0,0),\cdots
F_{10}=(0,0,0,0,\phi_{5})~.}
Note that with the choices made, the five $F^{i}$ are linearly
independent.

To complete our
assignment of transformation laws,
assign the $\lambda$s the following transformation properties under
$g_{1}$:
\eqn\glambdaone{g_{1}:~~(\lambda_{1},\lambda_{2},\lambda_{3},\lambda_{4},
\lambda_{5}) \rightarrow (\lambda_{2},\lambda_{3},\lambda_{4},\lambda_{5},
\lambda_{1})}
\eqn\glambdatwo{g_{1}:~~(\lambda_{6},\lambda_{7},\lambda_{8},\lambda_{9},
\lambda_{10}) \rightarrow (\lambda_{7},\lambda_{8},\lambda_{9},\lambda_{10},
\lambda_{6})~.}
If we now recall (see e.g. \DK) that
the $\lambda_{a}$ and the $F_{a}^{i}$ really enter the action of our
worldsheet field theory in the combination $P_{i}\lambda_{a}F_{a}^{i}$
where the $P_{i}$s are (0,2) chiral superfields, then we see that to truly
make the quantum field theory action invariant under $g_{1}$ we should also
have $g_{1}$ act to permute the $P_{i}$ in the same way as the $\phi_{i}$.
It is apparent that with this
choice of the $F$s and the action of $g_{1}$
on $V_{1}$, the $Z_{5}$ symmetry does indeed lift to an automorphism of
$V_{1}$.  Diagonalizing the action of $g_{1}$ on the $\phi$s, $P$s and
$\lambda$s, one can verify that the quadratic level-matching condition
\neven\ is satisfied by the eigenvalues of $g_{1}$.

Now that we have successfully taken the $g_{1}$ orbifold (and are down
to a 15 generation model), we must consider how to lift the action of
$g_{2}$ to $V_{1}$.  It is clear from the choice of the $F$s above
that we must assign the $\lambda$s the following transformation
under $g_{2}$ to keep $\lambda_{a}F_{a}^{i}$ invariant:
\eqn\gtwolone{g_{2} ~{\rm on}~ \lambda_{1,\cdots 5}:~~\alpha^{3},\alpha,
\alpha^{4},\alpha^{2},1}
\eqn\gtwoltwo{g_{2} ~{\rm on}~ \lambda_{6,\cdots,10}:~~\alpha^{4},
\alpha^{3},
\alpha^{2},\alpha,1~.}
And again it is simple to check that \neven\ is satisfied.

So finally, with
the data specified above, the full $Z_{5}\times Z_{5}$ symmetry
of the Fermat quintic lifts to an automorphism of $V_{1}$.  Taking the
$Z_{5}\times Z_{5}$ orbifold, we obtain a three generation $SU(5)$
theory, on a target manifold with $\pi_{1} = Z_{5}\times Z_{5}$.

\newsec{\bf {SU(5)~$\rightarrow$~SU(3)$\times$SU(2)$\times$U(1)}}

In compactification on a manifold $M$ with $\pi_{1}(M)\neq 0$,
one is allowed to give expectation values to Wilson lines
around the noncontractable loops $\gamma$ in $M$
\eqn\wilson{U_{\gamma}~=~{\rm P}~{\rm exp}~\left( \oint_{\gamma}
A ~dx \right)~.}
This amounts to
a choice of a homomorphism from $\pi_{1}(M) \rightarrow G$,
where $G$ is the spacetime gauge group.
One is left with a vacuum with $G$ broken to the subgroup of
$G$ which commutes with $U_{\gamma}$.

We will be choosing Wilson lines in some $Z_{N}$ subgroup of
the spacetime gauge group (corresponding to a homomorphism
mapping a $Z_{N}$ subgroup of $\pi_{1}(M)$ to $G$).
One can think of the Wilson lines as acting on the gauge degrees of
freedom of the heterotic string.
We are interested in using Wilson lines in the $SU(5)$ theories of
\S3, so we take the relevant worldsheet gauge degrees of freedom to
be four free left-moving bosons $X^{I}$.
Turning on Wilson lines means that we should include sectors in
which the $X^{I}$ (which live on a left-moving torus)
only close up to
\eqn\Wilsorb{X^{I}(\sigma + \pi) ~=~X^{I}(\sigma) + 2\pi \delta^{I}~}
for some $\delta^{I}$. According to \refs{\Wilmod,\dealwis}\
we should impose
the following analogue of the level-matching constraint \neven\ on this
orbifolding associated with the Wilson lines, if we are to obtain a
modular invariant theory.

As in \Segre, choose
\eqn\wilschoice{U~=~{\rm exp}\left( 2\pi i
\sum_{i=1}^{4}\delta^{i}
H^{i}\right)~}
where the $H^{i}$ generate the Cartan subalgebra of $SU(5)$.
Consider $(\delta^{1},\cdots,\delta^{4})$ as a four-vector in the dual
basis.
Then the level-matching condition for Wilson lines
is
\eqn\levelmat{{1\over 2} (\delta,\delta) ~=~0 ~{\rm mod}~{1\over N}~}
where the inner product $(\delta,\delta)~=~\delta_{i}A_{ij}\delta_{j}$
should be taken with an
insertion of the $A_{4}$ Cartan matrix $A_{ij}$.

Following p. 71 of \Slansky, we see that the unique choice (up to scale)
of $\delta$ which
breaks
$SU(5)$ to $SU(3)\times SU(2)\times U(1)$ (and corresponds to
choosing a Wilson line in the weak hypercharge $U(1)$)
is given by
\eqn\breaksm{\delta ~\sim~\left( -2,1,-1,2 \right) ~.}
So we need to find such Wilson lines which correspond to $Z_{3}$
transformations (for the case of Example 3) and $Z_{5}$ transformations
(for the case of Example 4) to accomplish the desired symmetry breaking.

In Example 3, with
$\pi_{1}(M) = Z_{3}$, take
\eqn\deltathree{\delta ~=~{1\over 3}(-2,1,-1,2)~.}
Some simple arithmetic tells us that
then $(\delta,\delta)~=~{10\over 3}$ which satisfies \levelmat.
In Example 4, with $\pi_{1}(M)=Z_{5}\times Z_{5}$, choose
\eqn\deltafour{\delta~=~{1\over 5}(-2,1,-1,2)~.}
This again satisfies \levelmat.
Turning on the Wilson lines \deltathree\ and \deltafour\ in Examples 3 and 4
of \S3, we find ourselves with two three generation models with gauge
group $SU(3)\times SU(2)\times U(1)$ at the string scale.

We have chosen to satisfy \levelmat\ separately
after constructing (0,2) models with $SU(5)$ gauge group in \S3 only
in order to make the physical interpretation (symmetry breaking by
Wilson lines on a manifold $M$ with $\pi_{1}(M)\neq 0$) manifest.
In general, one could orbifold by a $Z_{N}$ symmetry choosing
any action on both the internal and the gauge degrees of freedom,
as long as the combined system obeys the level-matching conditions.

\newsec{Conclusion}

In this paper we have discussed the construction of (0,2) Calabi-Yau
compactifications which can serve as suitable starting points for
string model building.
For (2,2) string models, one has well developed techniques to compute
the tree-level kinetic terms and the
tree-level superpotential (which receives no string loop corrections).
Similar progress in understanding the low energy effective actions of
more general (0,2) models is desirable.
In addition, the data on unification of couplings in
the minimal supersymmetric standard model provides a
challenge for string models which
yield $SU(3)\times SU(2)\times U(1)$ directly at the string scale
(see e.g. \threshold\ and references therein).
The computation of threshold corrections in general (0,2) models
might help in addressing this concern.

\bigskip
\centerline{\bf{Acknowledgements}}

I would like to thank J. Distler
for very helpful comments on a draft and
the authors of \WitSilvtwo\ for informing me of their
results prior to publication.  I would also like to thank Philip Candelas
for correcting a bad choice of data in Example 4 in an earlier
version of this paper.
This work was supported in part by a fellowship from the Harvard Society of
Fellows and by the William F. Milton Fund of Harvard University.

\listrefs
\end